# Simulations of scrape-off layer power width for EAST H-mode plasma and ITER 15 MA baseline scenario by 2D electrostatic turbulence code

X. Liu[1, *], A. H. Nielsen[2], J. J. Rasmussen[2], V. Naulin[2], L. Wang[1, 3], R. Ding[1], and J. Li[1, 3]

[1]*Institute of Plasma Physics, Chinese Academy of Sciences, Hefei 230031, People's Republic of China*

[2]*Department of Physics, Technical University of Denmark, Kongens Lyngby 2800, Denmark*

[3]*Institute of Energy, Hefei Comprehensive National Science Centre, Hefei 230031, People's Republic of China*

[*]E-mail: xliu@ipp.ac.cn

**Abstract**

The scrape-off layer power width ($\lambda_q$) is an important parameter for characterizing the divertor heat loads. Many experimental, theoretical, and numerical studies on $\lambda_q$ have been performed in recent years. In this paper, a 2D electrostatic turbulence code, BOUT-HESEL, has been upgraded to simulate H-mode plasmas for the first time. The code is validated against the previous implementation and the experimental $\lambda_q$ scalings for L-mode plasmas and experiments with a typical EAST H-mode discharge. The simulated $\lambda_q$ is found to agree quite well with the Eich scaling [Eich *et al.* 2013 *Nucl. Fusion* 53 093031] for the EAST H-mode discharge and the comparison of the probability distribution function of the parallel particle flux with the measurements by reciprocating probes is also consistent. The code is utilized to simulate the ITER 15 MA baseline scenario [Kim *et al.* 2018 *Nucl. Fusion* 58 056013]. The ITER simulation reveals that the radial particle/heat transports are dominated by blobby transports, and predicts $\lambda_{q,ITER}$ = 9.6 mm, which is much larger than the prediction by the Eich scaling ($\lambda_{q,ITER} \approx 1$ mm). Based on the EAST modified cases, an estimated HESEL H-mode scaling, $\lambda_q = 0.51 R_c^{1.1} B_t^{-0.3} q_{95}^{1.1}$ is proposed. This scaling predicts $\lambda_{q,ITER}$ = 9.3 mm, which agrees surprisingly well with that for the ITER case. A further investigation combined with the basic parameters in the database of the Eich scaling shows that the missing positive scaling dependence on the machine size ($R_c$) in the Eich scaling appears to be shaded by the negative scaling dependence on the toroidal magnetic field ($B_t$) for current devices. This is however not the case for ITER, explaining why simulations in recent studies and this paper can reproduce the Eich scaling for current devices, but predict a much larger $\lambda_q$ for ITER. According to the simulation results, the strong positive scaling dependence of $\lambda_q$ on $R_c$ is due to a combination of slowing down the parallel heat transports by increasing the parallel connection length and the enhancement of the radial E × B turbulent heat transports when the machine size is increased.

# 1. Introduction

For future magnetically confined devices, one of the crucial issues is to handle the excessive heat loads on the divertor targets [1]. In addition to applying active control methods, like divertor detachment by impurity seeding, understanding the physical mechanisms of how heat is transported in the edge and scrape-off layer (SOL) is also of great importance to predict and/or control the heat deposition on the divertor targets. The SOL power width ($\lambda_q$) is an important parameter to characterize the heat transports in the SOL and the heat deposition on the divertor targets. Several works [2-21] have been performed in recent years to study $\lambda_q$ and we only address the most relevant results in this paper. One of the most prominent experimental results is the multi-machine (or Eich) scaling [2,3]: $\lambda_q$ is nearly inversely proportional to the poloidal magnetic field ($B_p$) at the last closed flux surface (LCFS) and is independent of the machine size (extrapolation to the ITER H-mode baseline scenario gives $\lambda_{q,ITER} \approx$ 1mm). The Eich scaling can be well explained by the heuristic drift-based (HD) model [4], which gives $\lambda_q \approx 4T_{LCFS}a/(ZeB_pR_cc_s)$, where $T_{LCFS}$ is the plasma temperature near the LCFS, $a$ and $R_c$ are minor and major radii, and $c_s$ is the ion sound speed. The nearly inverse scaling dependence on the poloidal magnetic field or the plasma current can be reproduced by the 3D gyro-kinetic code (XGC1) [12], the 3D BOUT++ 6-field turbulence code [13,14], the SOLPS transport code [15], and the 2D BOUT-HESEL turbulence code [19] for current tokamaks. However, the simulations by XGC1 [12] and BOUT++ 6-field turbulence code [16,17] both predict a much larger $\lambda_q$ for ITER than the prediction by the Eich scaling. The possible reasons for this difference are: a) the Eich scaling misses a scaling dependence on the machine size (or this is cancelled out by some unknown scaling parameters) [12,18,19]; b) turbulences, rather than magnetic and/or E × B drifts for current devices, dominate the radial heat transport for ITER [16]. The key mechanism for the latter explanation is the magnetic drift, which indeed does not produce scaling dependence on the machine size and is the main mechanism for the drift-turbulence transition (radial heat diffusivity does not influence $\lambda_q$ when magnetic drift is dominant over turbulence) in the drift-turbulence competition theory [16], but its positive scaling dependence on the plasma edge temperature is against some experimental scalings [6,8]. Thus, further investigations are still in demand to fully understand the underlying physical mechanisms that determine $\lambda_q$.

In this paper, the 2D electrostatic turbulence code, BOUT-HESEL (implemented under the BOUT++ framework [22]), has been upgraded and can be used to simulate H-mode plasma parameters for the first time, making it possible to simulate ITER H-mode baseline scenario. BOUT-HESEL utilizes the 2D slab geometry with parameterized parallel dynamics modelled as loss terms in the SOL and drift wave terms in the edge region, which naturally are limitations compared with 3D codes like XGC1 and BOUT++ 6-field turbulence code. But this simplification allows the simulations for running much longer time series of the evolutions to provide better statistics and indeed also performing faster and wider parameter scans. The 2D slab geometry does not include the divertor region, the high-field side, and the toroidal direction. Thus, full parallel dynamics, ballooning effects, and divertor dynamics are missing in BOUT-HESEL. This simplified geometry also makes it impossible to study the influence of plasma shape (like elongation and triangularity) on the simulation results (herein $\lambda_q$). Since BOUT-HESEL is an electrostatic code, it is not capable of describing electromagnetic dynamics, which is acceptable in simulations for plasmas with low and intermediate values of beta (the effect of electromagnetic perturbations on the SOL turbulences is negligible) [21]. Unlike BOUT++ 6-field turbulence code, the lack of magnetic perturbations and ballooning effects limits BOUT-HESEL in the simulation of ELMy bursts for ELMy H-mode plasmas, which is not a big issue in the simulation of H-mode plasma during inter-ELM phase. The fluid description of the HESEL model also misses the kinetic effects in low-collisional plasmas compared with XGC1. Recent study shows that kinetic effects may challenge the fluid description of blob dynamics in the parallel direction [23]. This issue will be further considered by feeding the parameterized parallel dynamics from extra 1D model with kinetic correction in future work. The last limitation we want to address is that the total power entered into the SOL ($P_{SOL}$) is highly dependent on the simulated final profiles, which normally differ from the initial profiles. This leads to a difference of more than one order of magnitude in $P_{SOL}$ for the ITER simulation. If an experimental result can be explained by the simpler HESEL model, then the dominant physics is probably included in this model. Having this in mind, the upgraded BOUT-HESEL can provide more information on the prediction of ITER $\lambda_q$ and the understanding of the underlying physics of $\lambda_q$. The rest of this paper is arranged as follows: Section 2 introduces the updated HESEL model and the benchmark results with the previous version and the experimental $\lambda_q$ scalings on simulating L-mode $\lambda_q$; section 3 validates the new code against experiments and studies the model with a typical EAST H-mode discharge; section 4 simulates the ITER 15 MA baseline scenario (Q = 10) and compares the simulation results with the previous studies; section 5 discusses the extrapolation of $\lambda_q$ from current devices to ITER and proposes an explanation for the prediction difference of $\lambda_q$ between the Eich scaling and the simulations; section 6 summaries the results of the paper.

## 2. The Reformulated HESEL model

The HESEL model is based on the Braginskii two-fluid equations with full collisions retained, aiming to simulate the electrostatic turbulence driven by the interchange instabilities in the edge and SOL [24,25]. The model has been successfully validated against experiments for L-mode plasmas in two different implementations [19,26]. In this paper, the HESEL model has been reformulated for $Z \neq 1$ and H-mode related parameters. Employing the gyro-Bohm normalization as described in Ref. [19], the model writes as

$$\partial_t n + \frac{1}{B}\{\varphi, n\} + n\kappa(\varphi) - \kappa(p_e) = \Lambda_n,$$
$$\partial_t w^* + \{\varphi^*, w^*\} - \kappa(p_i + p_e) = \Lambda_{w^*},$$
$$\frac{3}{2}\partial_t p_e + \frac{3}{2}\frac{1}{B}\{\varphi, p_e\} + \frac{5}{2}p_e\kappa(\varphi) - \frac{5}{2}\kappa(T_e p_e) = \Lambda_{pe}, \quad (1)$$
$$\frac{3}{2}\partial_t p_i + \frac{3}{2}\frac{1}{B}\{\varphi, p_i\} + \frac{5}{2}p_i\kappa(\varphi) + \frac{5}{2}\frac{1}{Z}\kappa(T_i p_i) - p_i\kappa(p_i + p_e) = \Lambda_{pi}.$$

The damping and drift wave terms arranged on the right-hand side of Eq. (1) are

$$\Lambda_n = -\nabla \cdot (n\vec{u}_R) - \sigma_{open}\frac{n}{\tau_n} - \sigma_{closed}\Im + \sigma_{force}\frac{n_{init} - n}{\tau_f},$$

$$\Lambda_{w^*} = \frac{3}{10}\frac{1}{Z}D_{i0}\nabla_\perp^2 w^* - \sigma_{open}\frac{w^*}{\tau_n} + \sigma_{open}S - \sigma_{closed}\Im,$$

$$\Lambda_{pe} = -\left[\nabla \cdot (p_e\vec{u}_R) - \left(\epsilon_{\kappa\perp} - \frac{9}{4}\right)\nabla \cdot (nD_e\nabla_\perp T_e) + \vec{u}_R \cdot \nabla p_i + Q_\Delta\right]$$
$$-\sigma_{open}\left(\frac{9}{2}\frac{p_e}{\tau_n} + \frac{\rho_{s0}}{L_c}q_{cond,e}\right) - \sigma_{closed}\left(\frac{5}{2} + \epsilon_{RT}\right)\bar{T}_e\Im + \sigma_{force}\frac{p_{e,init} - p_e}{\tau_f}, \quad (2)$$

$$\Lambda_{pi} = -\left[\frac{5}{2}\nabla \cdot (p_i\vec{u}_R) - \frac{2}{Z}\nabla \cdot (nD_i\nabla_\perp T_i) - \vec{u}_R \cdot \nabla p_i - Q_\Delta\right]$$
$$-\sigma_{open}(\frac{9}{2}\frac{p_i}{\tau_n} + \frac{\rho_{s0}}{L_c}q_{cond,i}) + \sigma_{open}p_iS - \sigma_{closed}\bar{p}_i\Im$$
$$+\frac{3}{10}\frac{n}{Z}D_i\left[4(\partial_{xy}^2\varphi^*)^2 + (\partial_x^2\varphi^* - \partial_y^2\varphi^*)^2\right] + \sigma_{force}\frac{p_{i,init} - p_i}{\tau_f},$$

where $\sigma_{open} = \frac{1}{2}\left[1 + tanh\left(\frac{x-x_{LCFS}}{\delta_{open}}\right)\right]$, $\sigma_{closed} = \frac{1}{2}\left[1 - tanh\left(\frac{x-x_{LCFS}}{\delta_{closed}}\right)\right]$ and $\sigma_{force} = \frac{1}{2}\left[1 - tanh\left(\frac{x-x_{force}}{\delta_{force}}\right)\right]$ are functions for domain separation (see figure 1 in Ref. [19]). As a continuous work, only the terms or symbols that have been added or modified in Eq. (1) and Eq. (2) compared with the previous formulation in Ref. [19] will be addressed in this section, for simplicity. The new model keeps full collisional effects and the resistive drift velocity changes to

$$\vec{u}_R = -\frac{D_e}{T_e}\left[\left(\frac{T_i}{Z} + T_e\right)\frac{\nabla_\perp n}{n} + \frac{\nabla_\perp T_i}{Z} - \frac{\nabla_\perp T_e}{2}\right]. \quad (3)$$

The elastic electron-ion collisional term that exchanges energy has the following form (there is a typo in Ref. [19])

$$Q_\Delta = 3\frac{m_e}{m_i}\frac{\nu_{ei0}}{\omega_{ci0}}n^2 T_e^{-3/2}(T_e - T_i). \quad (4)$$

The drift wave term uses a slightly different form

$$\Im = \frac{1}{\tau_{DW}}(\frac{\bar{T}_e}{\bar{n}}\tilde{n} + \tilde{T}_e - \tilde{\varphi}). \quad (5)$$

For L-mode high-collisional plasmas, the Spitzer-Härm conduction is adequate to represent the electron and ion parallel conduction in the SOL

$$q_{cond,e} = q_{sh,e} = \frac{L_c}{\rho_{s0}} \frac{1}{\tau_{she}} T_e^{\frac{7}{2}},$$

$$q_{cond,i} = q_{sh,i} = \frac{L_c}{\rho_{s0}} \frac{1}{\tau_{shi}} \frac{T_{e0}^{5/2}}{T_{i0}^{5/2}} T_i^{7/2}.$$

(6)

While for H-mode low-collisional plasmas, the flux-limited conduction

$$q_{fl,e} = \alpha_{fl,e} n (m_i/m_e)^{1/2} T_e^{3/2},$$

$$q_{fl,i} = \alpha_{fl,i} \frac{n}{Z} T_i^{3/2},$$

(7)

are typically conjunct with the Spitzer-Härm conduction to characterize the parallel conduction [27]. $\alpha_{fl,e}$ and $\alpha_{fl,i}$ are free constants and are related to the electron and ion collisionalities, respectively. As $\alpha_{fl,e}$ or $\alpha_{fl,i}$ decreases, the contribution from the flux-limited conduction to the parallel conduction increases. The conjunct conduction bridge the high-collisional plasma in the far SOL and the low-collisional plasma in the near SOL, which write as [27]

$$q_{cond,e} = \frac{1}{1/q_{fl,e} + 1/q_{sh,e}},$$

$$q_{cond,i} = \frac{1}{1/q_{fl,i} + 1/q_{sh,i}}.$$

(8)

The characteristic times in the parallel dynamics are modified to be

$$\tau_n = \frac{L_b}{2Mc_s} \omega_{ci0},$$

$$\tau_{she} = \frac{L_c^2 m_e \nu_{ei0}}{\epsilon_{\kappa//} T_{e0}} \omega_{ci0},$$

$$\tau_{shi} = \frac{L_c^2 Z m_i \nu_{ii0}}{3.9 T_{i0}} \omega_{ci0},$$

$$\tau_{DW} = \frac{L_c^2 \epsilon_{Ru} m_e \nu_{ei0}}{T_{e0}} \omega_{ci0},$$

(9)

where $\omega_{ci0} = \frac{eZB_0}{m_i}$, and $c_s = \sqrt{\frac{ZT_e + T_i}{m_i}}$. The values of the constants in the collisional terms are related to the charge number of main ion species, as listed in table 1.

Compared with the previous model in Ref. [19], the formulation of the updated model does not change significantly. The new model is reduced to the old one by simply setting Z = 1. Although the new update enables the study of $\lambda_q$ for ions with Z > 1, we focus on simulating H-mode plasmas in this paper. Some primary test cases that compare deuterium and helium plasmas with the same pressure profiles show that $\lambda_q$ has no significant difference. A detailed study on this topic will be carried out in future work. Other than this, the main difference is that the new model keeps the ion diamagnetic drift velocity in the advective part of the gyro-viscous and polarization term (second term of vorticity equation in Eq. (1)) after carrying out the gyro-viscous cancellation. This change does not influence the simulation of $\lambda_q$ significantly according to some test cases based on the EAST L-mode discharge.

Table 1. Values of the constants in the collisional terms for different charge numbers [25].

| Z | $\epsilon_{Ru}$ | $\epsilon_{RT}$ | $\epsilon_{\kappa//}$ | $\epsilon_{\kappa\perp}$ |
|---|---|---|---|---|
| 1 | 0.51 | 0.71 | 3.16 | 4.66 |
| 2 | 0.44 | 0.9 | 4.9 | 4.0 |
| 3 | 0.40 | 1.0 | 6.1 | 3.7 |
| 4 | 0.38 | 1.1 | 6.9 | 3.6 |
| ∞ | 0.29 | 1.5 | 12.5 | 3.2 |

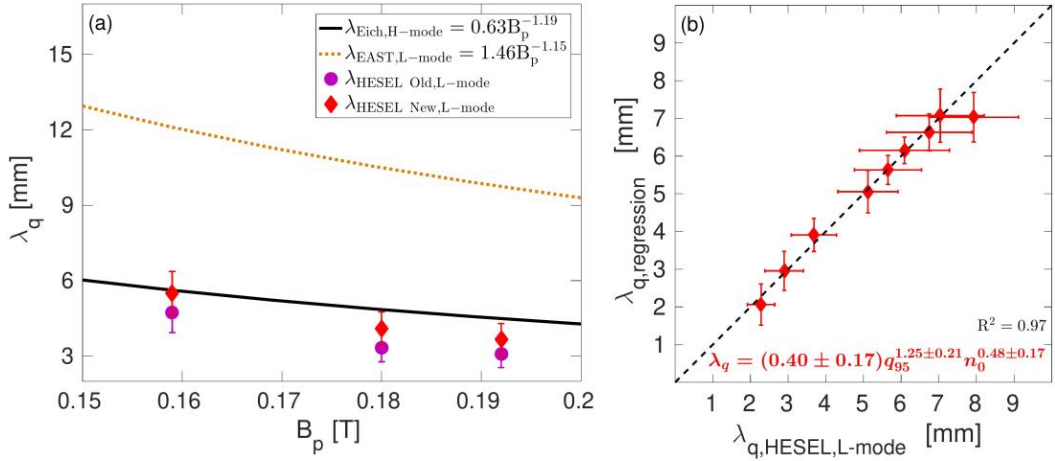

Figure 1 (a) Comparisons of the simulated $\lambda_q$ with the experimental scalings. The black solid line and the orange dotted line represent the Eich H-mode and EAST L-mode scalings, respectively. (b) The numerical L-mode scaling of $\lambda_q$ with respect to the edge safety factor ($q_{95}$) and the density at the separatrix of the initial profile ($n_0$).

Three L-mode discharges in Ref. [19] have been simulated again to firstly validate the new model. The simulation setups are kept almost unchanged. Figure 1(a) compares the simulated $\lambda_q$ with the experimental scalings. The new model produces more consistent results with the Eich scaling compared with the old ones. However, the differences of the simulated $\lambda_q$ between these two versions are not significant, meeting our expectations. Recently, a multi-machine L-mode scaling was proposed to predict ITER L-mode $\lambda_q$, i.e., $\lambda_{q,Horacek} = 2.8 \times 10^3 (a/R_c)^{1.03} f_{GW}^{0.48} j_p^{-0.35}$, where $f_{GW}$ is the Greenwald density fraction and $j_p$ is the plasma current divided by the area of the plasma cross-section [11]. According to the results shown in figure 3 of Ref. [11], the most significant scaling parameters are the line-averaged electron density $\bar{n}_e$ and the well-known scaling parameter $B_p$. Then this L-mode scaling can be rewritten as $\lambda_{q,Horacek} = 2.8 \times 10^{5.4} \mu_0^{0.83} (2+2\kappa^2)^{-0.415} \kappa^{0.35} a^{1.86} R_c^{-1.03} \bar{n}_e^{0.48} B_p^{-0.83} \propto \bar{n}_e^{0.48} B_p^{-0.83}$. Based on the EAST L-mode discharge, the edge safety factor ($q_{95}$) and the initial density profile (the value at the separatrix $n_0$ is used as the scaling parameter) are scanned and a numerical $\lambda_q$ scaling ($\lambda_{q,HESEL,L-mode} = 0.40 n_0^{0.48} q_{95}^{1.25}$) is obtained as shown in figure 1(b). This numerical scaling shows excellent agreement with the experimental multi-machine L-mode scaling in the scaling dependence on the electron density ($n_0 \propto \bar{n}_e$) and good agreement with the Eich scaling ($\lambda_{q,Eich} = 0.63 B_p^{-1.19}$) in the scaling dependence on $B_p$ ($q_{95} \propto 1/B_p$). It is also consistent with the theoretical scaling of L-mode pressure gradient length ($L_p \propto n_e^{10/17} q_{95}^{12/17}$, see Eq. (29) in Ref. [21]), if $\lambda_q \propto L_p$ is assumed. These comparison results confirm that the reformulated HESEL model is adequate for simulating L-mode $\lambda_q$.

## 3. Simulation of EAST H-mode plasma

One of the main goals to update the HESEL model is to study the SOL power width for H-mode plasmas, which has not been performed in the previous work due to the lack of enough stabilizing mechanism to handle the high-k instabilities in the numerical scheme with finite grid resolution for low-collisional plasmas. In this paper, hyper-viscosity is introduced to overcome this difficulty. As will be described in this section, low-level hyper-viscosity does not change $\lambda_q$ significantly, making it possible to simulate the ITER 15 MA baseline scenario in section 4.

### 3.1 Simulation setup

A typical ELMy H-mode discharge in EAST with a lower single null (LSN) configuration is selected for our simulations. The plasma was heated by lower hybrid wave (LHW, $P_{LHW}$ = 0.46 MW), electron cyclotron heating (ECH, $P_{ECH}$ = 0.95 MW), and neutral beam injection (NBI, $P_{NBI}$ = 1.30 MW). The other main parameters are: toroidal magnetic field $B_t$ = 2.44 T, plasma current $I_p$ = 500 kA, safety factor at 95% flux surface $q_{95}$ = 5.53, and plasma stored energy $W_{MHD}$ = 184.7 kJ. The wall was conditioned with lithium and the plasma was fueled by super-sonic molecule beam injection.

The initial profiles used for the simulation are shown in figure 2. These profiles are obtained similarly as in Ref. [19], where the diagnostics employed for the measurements and the fitting method are described in detail. In this discharge, the ion temperature is lower than the electron temperature in the center, which might result from the high heating efficiency by ECH. The simulation region is a rectangle (292 mm long in the radial direction and 146 mm wide in the poloidal direction), corresponding to a 576 × 288 box in grids. The grid spacing is $dx = 0.76\rho_{s0}$, where $\rho_{s0}$ = 0.66 mm. Given the reference values of the plasma density and temperatures at the separatrix ($n_0 = 1.84 \times 10^{19}$m$^{-3}$, $T_{e0} \approx T_{i0} \approx$ 82 eV), the normalized SOL collisionalities for electron and ion are, $\nu_e^* \equiv \frac{L_c \nu_{ei0}}{v_{the0}} \approx 2.9$ and $\nu_i^* \equiv \frac{L_c \nu_{ii0}}{v_{thi0}} \approx 2.0$. Although $\nu_e^*$ and $\nu_i^*$ are greater than unity, it requires higher collisionality for Spitzer-Härm conduction to fully hold [27]. So, the conjunction of Spitzer-Härm conduction with flux-limited conduction, i.e., Eq. (8), is chosen to better describe the parallel conduction in the SOL. The coefficients of the electron and ion flux-limited conduction are assumed to be the same, i.e., $\alpha_{fl,e} = \alpha_{fl,i} = \alpha$. We set $\alpha = 0.3$ [27] as the default value in this section for the numerical study of the HESEL model.

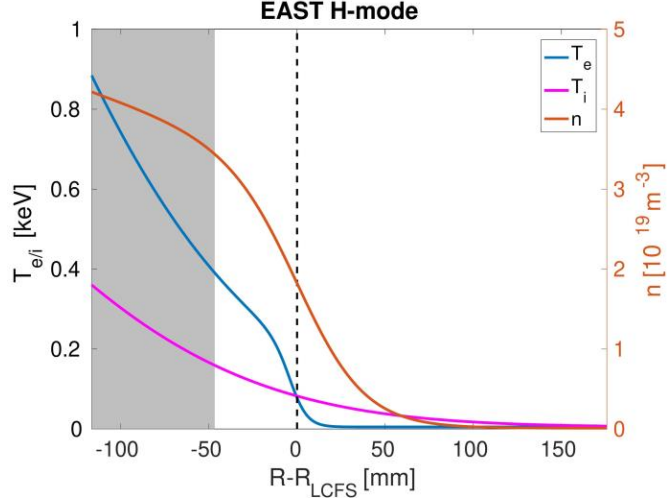

Figure 2 The initial profiles used for the EAST simulation. These profiles are obtained by fitting the experimental measurements in the edge and near SOL and the background paddings in the far SOL. The dashed vertical line represents the separatrix. The shaded area represents the profile forcing region ($R_{PFR}$ = 0.6).

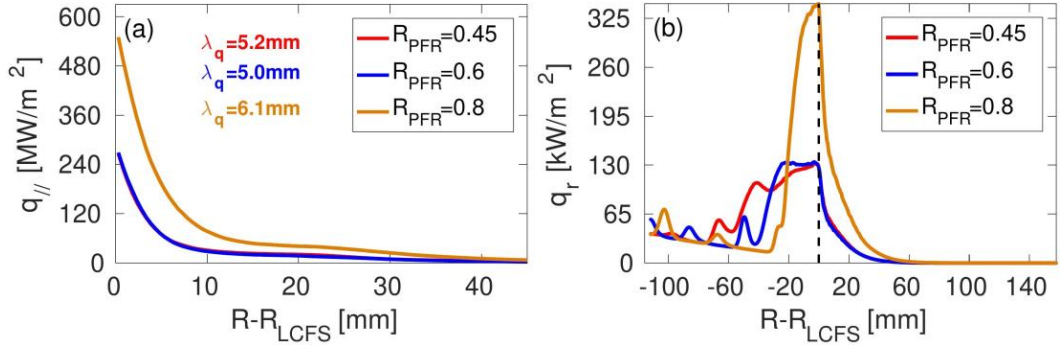

Figure 3 The profiles of the (a) total parallel heat fluxes in the SOL and (b) total radial heat fluxes for different widths of the profile forcing region (PFR). The dashed vertical line represents the separatrix.

In the previous work [19], the width of the profile forcing region (PFR), where the plasma profiles are forced toward the specified initial profiles within a given relaxation time (see Eq. (2)) to sustain the pressure gradient, occupied half of the edge region. The reason for introducing a PFR region as the soft boundary condition instead of using a fixed inner boundary condition for the plasma density and electron and ion temperature fields is to keep the gradient of the evolving pressure profile in the PFR that drives the interchange instability close to the experimental one and avoid the 'unphysical' gradients near the radial inner boundary when applying Dirichlet boundary conditions with fixed values. The presence of PFR somehow shrinks the actual simulation domain size in the radial direction. In this sense, the actual radial domain size changes along with the width of PFR. Since the gradient of the initial pressure profile normally varies radially, the width of PFR (so as the actual radial domain size) may be important to the simulation results. The ratio of the width of PFR to the width of the edge region, $R_{PFR}$, is scanned to evaluate the influence of PFR on $\lambda_q$, and the results are shown in figure 3. A large $R_{PFR}$ ($R_{PFR}$ = 0.8) has a significant influence on both the parallel and radial heat fluxes resulting from not enough space in the edge region for turbulent fluctuations, and a medium $R_{PFR}$ ($R_{PFR}$ = 0.45 and 0.6) seems to have no significant influence on the parallel heat flux in the SOL. The small bumps of the radial heat fluxes in the PFR result from a relatively loose setting for profile forcing. Since the radial heat flux near the separatrix is flat (due to energy conservation) for $R_{PFR}$ = 0.6, this value is set for the rest of EAST simulations in this section.

### 3.2 The influence of hyper-viscosity on SOL power width

The free energy released from the pressure gradient is transferred to turbulent energy (includes low-k and high-k components), which is finally dissipated to heat by diffusions if there are no sinks. In HESEL simulations, the high-k numerical components or instabilities could not be efficiently dissipated or stabilized if the collisionality is too small in numerical scheme with finite grid resolution, making the simulations fail to converge. To suppress the harmful high-k numerical instabilities, the hyper-viscosity (applied only to the edge region by multiplying with the function $\sigma_{closed}$ as defined in the text below Eq. (2)) is added into the density and vorticity equations

$$\begin{aligned} \Lambda_n &= \cdots - \sigma_{closed}\beta D_{e0}\nabla_\perp^4 n, \\ \Lambda_{w^*} &= \cdots - \sigma_{closed}\beta \frac{3}{10}\frac{1}{Z}D_{i0}\nabla_\perp^4 w^*. \end{aligned} \quad (10)$$

The coefficients of the hyper-viscosity are $\beta$ times the coefficients of the classical diffusion. Similar to diffusion terms, the introduced hyper-viscosity terms also damp the turbulent energy to heat, but their damping strengths are much smaller and mainly affect the high-k components. So, the inclusion of hyper-viscosity in the simulation will somehow weaken the turbulences. For the EAST (see figure 2) and ITER (see figure 9) initial density profiles, the minimum value of the radial characteristic length near the separatrix in the edge region has an order of $100\rho_{s0}$, meaning that the hyper-viscosity is at least $10000/\beta$ times smaller than the classical diffusion. So, the introduced hyper-viscosity terms should have small influence on the simulation results with small value of $\beta$.

In figure 4, $\beta$ is scanned to validate the functionality of the hyper-viscosity. From the calculated power spectrum of the poloidal distribution of the density at $R$-$R_{LCFS}$ ≈ -23 mm in figure 4(a), we see that: i) the power intensity decreases steadily in the low-k regime and drops dramatically in the high-k regime; ii) the larger the $\beta$, the lower the power intensity in the high-k regime. Figure 4(b) shows that the total parallel heat flux decreases when $β$ increases. Although $λ_q$ also has a decreasing trend, it does not change significantly for low-level $β$. The hyper-viscosity is deliberately introduced into the HESEL model to affect only the particle transports in the edge region, aiming to suppress the numerical high-k instabilities (spatial scales approaching the grid resolution) without changing the SOL heat transports significantly, which is demonstrated to be effective for low-level $β$.

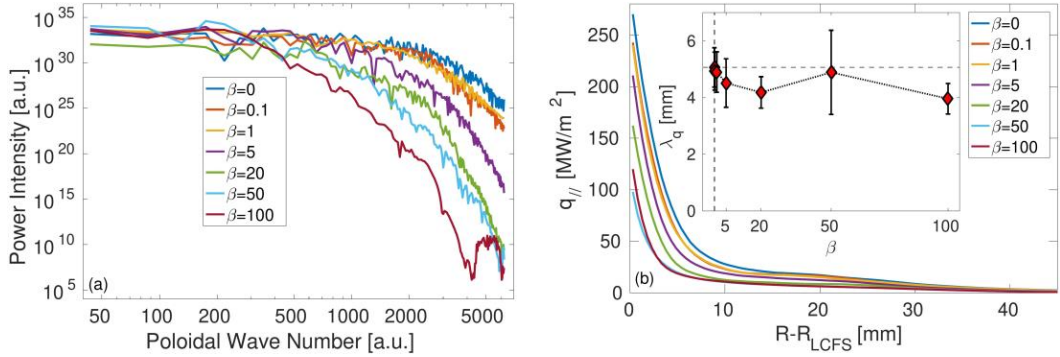

Figure 4 (a) The power spectrum of the poloidal density distribution at $R$-$R_{LCFS}$ ≈ -23 mm for different levels of hyper-viscosity (represented by $β$). (b) The radial distribution of the total parallel heat flux for different $β$. The inserted picture plots $λ_q$ versus $β$.

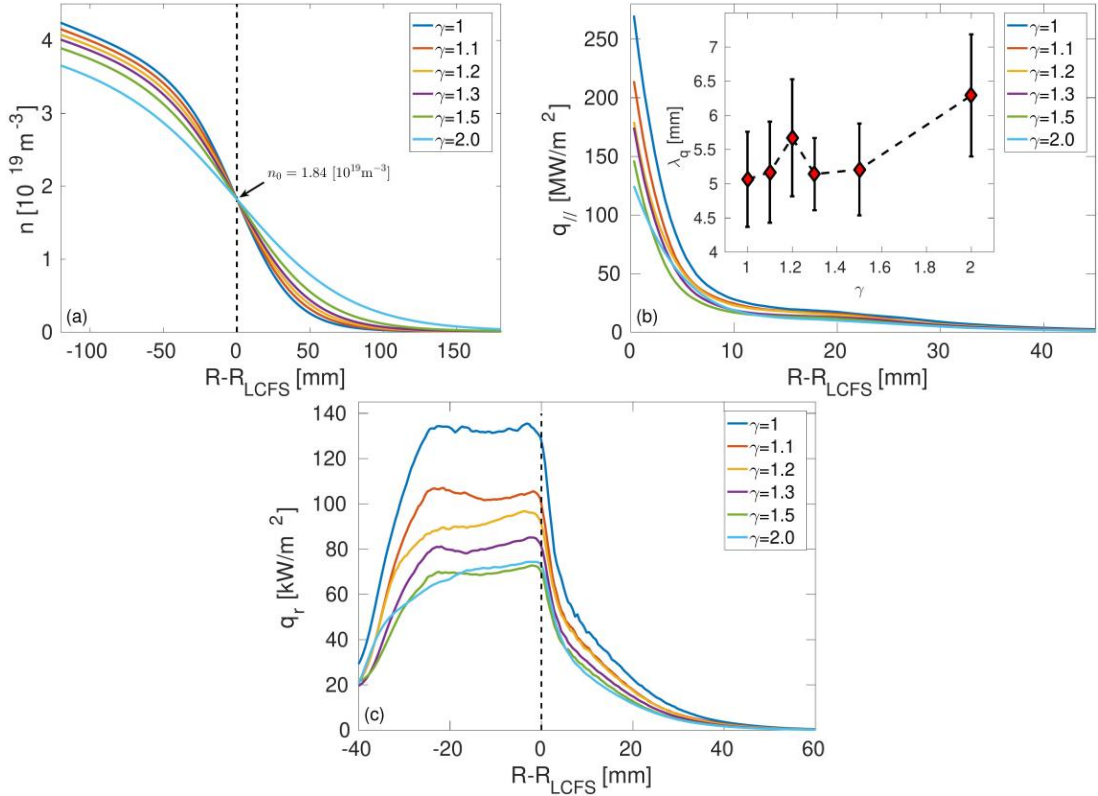

Figure 5 (a) Demonstration of stretched initial profiles for the density profile, where $γ$ is the stretching factor ($γ$ = 1 represents the original profile). (b) The radial distributions of the total parallel heat fluxes for different γ. The inserted picture plots $\lambda_q$ versus γ. (c) The radial distributions of the total radial heat fluxes for different γ. The dashed vertical lines in (a) and (c) represent the separatrix.

### 3.3 The influence of pressure gradient on SOL power width

As mentioned above, pressure gradient influences interchange instabilities in the edge region. To further evaluate its influence on $\lambda_q$, the initial profiles in figure 2 are stretched with a factor of γ to change the pressure gradient. Since the values of the initial profiles at the LCFS are important input parameters for BOUT-HESEL, they are kept unchanged using the following stretching method: change the values of the horizontal axis from $R$-$R_{LCFS}$ to γ ($R$-$R_{LCFS}$) and keep the values of the vertical axis fixed for all the profiles in figure 2. Figure 5(a) shows an example of the stretched initial profiles with different γ (γ = 1 is the original profile) for the density profile (the electron and ion temperature profiles are stretched in the same way). With the stretched initial profiles, the simulations are performed using the same simulation domain and setups. Figure 5(b) and 5(c) show the distributions of the total parallel and radial heat fluxes, respectively. As γ increases, both the parallel and radial heat fluxes decrease, but $\lambda_q$ does not change significantly for γ ≤ 1.5 (see the inserted picture in figure 5(b)). This indicates that $\lambda_q$ is possibly determined by the dynamics in the SOL and is highly dependent on the plasma parameters at the separatrix (see the L-mode scalings in Refs. [18,19]). For low-level γ, the way the dynamics in the edge region (or herein the pressure gradient) could influence $\lambda_q$ significantly appears to be through changing the plasma parameters at the separatrix.

### 3.4 Comparisons with experiments

The inclusion of flux-limited conduction for the low-collisional SOL introduces a free parameter $\alpha$, which significantly influences the SOL power width, since $\alpha$ governs the mechanism of the parallel conduction (see Eqs. (6-8)). The value of $\alpha$ is related to the SOL collisionality. For a collisionless SOL, a smaller $\alpha$ is expected to count the kinetic modification of the parallel conduction. However, there is a lower limit for $\alpha$ (0.08 for the electron in the EAST simulations), below which 'unphysical' effects may become a problem [28]. In this subsection, $\alpha$ is scanned from 0.15 to 1.2 for the EAST H-mode discharge and its exact value is determined by comparison with the experimental results.

Figure 6 shows the detailed distributions of the parallel and radial heat fluxes for the scan of $\alpha$. The dominant composition of the parallel heat flux in the near SOL is the electron heat conduction, while the ion heat conduction is negligible in the whole SOL region. In the far SOL, the ion heat advection becomes important. As $\alpha$ increases, the parallel electron heat conduction increases dramatically in the near SOL due to the increased contribution from the Spitzer-Härm conduction, which is more sensitive to the electron temperature, while the electron and ion heat advections do not change significantly (advections are mainly determined by the less affected ion temperature and plasma density). So, no surprise that $\lambda_q$ (mainly determined by the distribution of the parallel heat flux in the near SOL) decreases with increasing $\alpha$. The radial heat fluxes of electrons and ions are dominated by the E × B turbulent transports in both the outer edge and SOL regions. Different from the parallel heat flux in the SOL, the radial heat flux in the outer edge region (about 120 kW/m$^2$) is not sensitive to $\alpha$, since $\alpha$ changes how energy is dissipated and distributed in the SOL, while the energy itself is generated in the edge region. Assume that particles and energy are expelled to the SOL mainly in the so-called ballooning region around the outboard midplane with a 60° poloidal extend [24,29]. Then we can estimate $P_{SOL}$ by

$$P_{SOL} \approx 2\pi(R_c + a)\frac{2\pi a}{6}q_{r,LCFS}. \quad (11)$$

The estimated value of $P_{SOL}$ for the EAST simulations is about 0.8 MW, which is in the same order of magnitude as the total input heating power ($P_{tot} \approx 2.7$ MW).

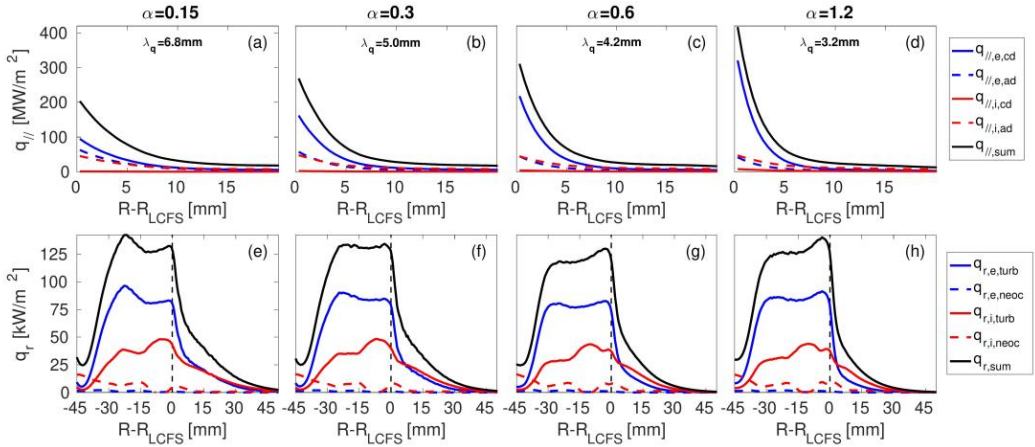

Figure 6 The distributions of the parallel and radial heat fluxes for different levels of flux-limited conduction (represented by $\alpha$). The red solid and dashed lines represent the heat fluxes for electron, the blue ones represent the heat fluxes for ion, and the black solid lines represent the sum of the parallel or radial heat fluxes. The solid lines and dashed lines represent conduction (cd) and advection (ad) for the parallel heat fluxes, respectively, and represent E × B turbulent (turb) and neo-classical (neoc) transports for the radial heat fluxes, respectively.

To determine the value of $\alpha$ ($\alpha$ influences the parallel particle flux in the SOL through changing the temperature profiles, which affect the parallel damping velocity of particles), the normalized probability distribution function (PDF) of the parallel particle flux ($\Gamma_{//}$) is compared with the measurements by the reciprocating probes [30] located at the outboard midplane (OMP). The measurement position of $\Gamma_{//}$ is $R$-$R_{LCFS} \approx 50$ mm and only the inter-ELM data are retained. $\Gamma_{//}$ sampled in different radial locations in the SOL of the EAST simulations for different $\alpha$ are compared with the experimental measurements in figure 7(a). The inserted picture shows the coefficient of determination ($R^2$) between the simulated PDFs with different sampling locations and the measured PDF with respect to $\alpha$. For the sampling location with $R$-$R_{LCFS} = 10$ mm, $R^2$ has the largest values for $\alpha = 0.3$ and 0.6. Although $R^2$ for $\alpha = 0.3$ has the largest value, $\alpha = 0.6$ is finally selected since the final electron temperature profile is closer to the initial profile in the SOL (especially their values at the LCFS are closer) as shown in the inserted picture of figure 7(b). The main frame of figure 7(a) plots the simulated and measured PDF × σ with respect to $\Gamma_{//}/\sigma$, which shows relatively good agreement. The difference of the radial location where $\Gamma_{//}$ is sampled or measured might result from the fact that the particles in the simulation are lost much faster in the SOL as compared with the experiment. Figure 7(b) shows the density profiles for the simulation and the measurement by the reciprocating probes. Note that here the density is measured by the triple probes and evaluated by the classical probe theory. The simulation profile is lower than the experimental profile and the densities at the sampling location ($R-R_{LCFS}$ = 10 mm) and the measurement location ($R-R_{LCFS} \approx 50$ mm) of $\Gamma_{//}$ are nearly the same, indicating that there might miss some mechanisms in the simulation to sustain the initial density profile. It should be clarified that unlike the concave tunnel probe, the convex probes that are used in the reciprocating probe system suffer from the sheath expansion effect and the erosion of the tip head, leading to uncertainty of evaluating the collecting area [31]. These may lead to measurement error of the ion saturation current (and thus the plasma density), which would further change the measured PDF. So, it might not be a good choice to use reciprocating probes as the best calibration tool for determining $\alpha$, especially for measurements in the H-mode plasmas. In BOUT++ 6-field turbulence code, $\alpha$ is determined by comparing the energy loss with experiments during the ELMy crash, which cannot be done for BOUT-HESEL. An ultimate solution for this issue is to replace current fluid description of the parallel dynamics with the kinetic description. A promising way is to combine the 1D parallel kinetic model, e.g., BIT1 [23], with the HESEL model. This will not only avoid the calibration of $\alpha$, but also gives a more consistent simulation of the SOL heat fluxes and a better prediction of $\lambda_q$. This work will be carried out in the future.

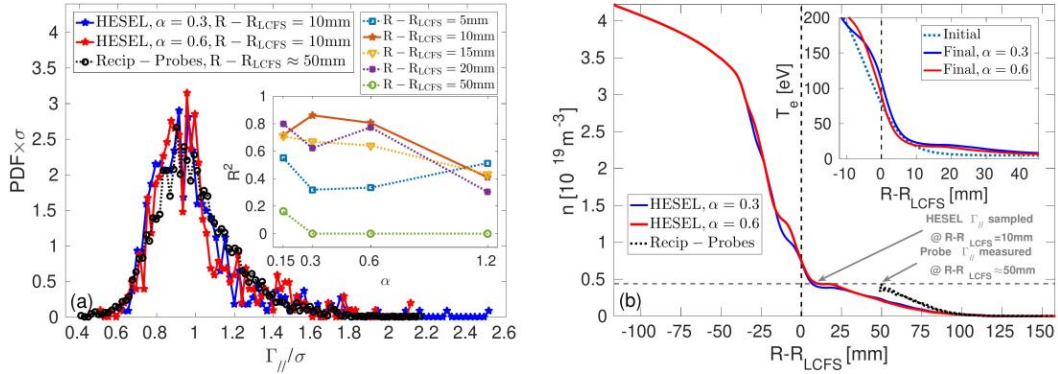

Figure 7 (a) Comparison of probability distribution function (PDF) of parallel particle flux ($\Gamma_{//}$) between the HESEL simulations and the experimental data measured by midplane reciprocating probes (Recip-Probes) for the EAST H-mode discharge. The inserted picture shows the coefficient of determination ($R^2$) between the simulated PDF × σ with different sampling locations and the measured PDF × σ for different $\alpha$. (b) The radial density distributions for the HESEL simulation and the Recip-Probes measurements. The inserted picture compares the final electron temperature profiles for $\alpha$ = 0.3 and 0.6 with the initial one. The text arrows show the radial locations where $\Gamma_{//}$ is sampled or measured in (a).

Figure 8 compares the simulated $\lambda_q$ with the experimental scalings and the HD model. Except for the EAST H-mode scaling, the simulated $\lambda_q$ for $\alpha = 0.6$ is almost the same as predicted by the Eich scaling and the HD model. The reason why the EAST scaling is almost two times larger than the Eich scaling is probably due to the radio-frequency heating scheme [32] or the different wall conditions compared with the other machines. So, it is reasonable that the simulated $\lambda_q$ agrees better with the Eich scaling than the EAST scaling, considering that the unknown physics that broadens $\lambda_q$ in the experiment might be missing in the HESEL model. The consistency between the HESEL simulation and the HD model will be further discussed in section 5. After all, the newly updated HESEL model can produce reasonable simulation results that are relatively consistent with the experiments for the selected EAST H-mode discharge, laying the foundation for simulating the ITER 15 MA baseline scenario.

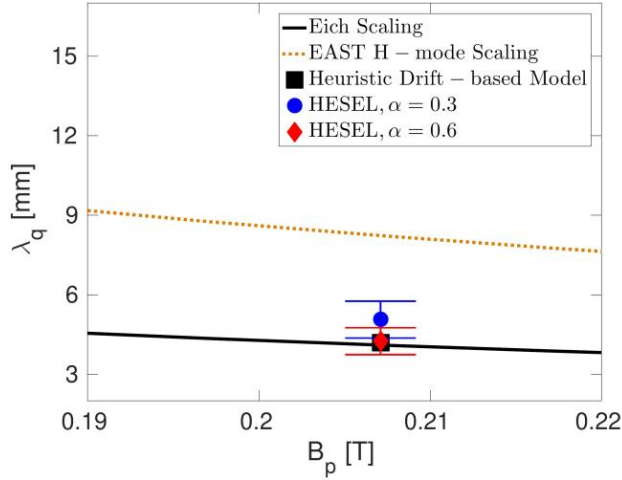

Figure 8 Comparisons of the simulated $\lambda_q$ with the experimental scalings and the heuristic drift-based model.

## 4. Simulation of ITER 15 MA Baseline Scenario

In the previous works of simulating the SOL power width for ITER by XGC1 [12] and BOUT++ 6-field turbulence code [16,17], the 15 MA baseline scenario (Q = 10) was chosen. For comparisons of $\lambda_q$ with these two codes, we choose the same scenario and use the initial profiles developed by Kim *et al.* [33]. Since the gradients of the original initial profiles are very steep in the edge region, the present simulation fails to converge in the very early phase (the steep profiles may not be effectively resolved by the relatively sparse grids in the simulation). The initial profiles are stretched with $\gamma = 1.2$ (see the dashed lines in figure 9) and the hyper-viscosity is turned on with $\beta = 5$ to solve the convergency problem. Note that the values of $\beta$ and $\gamma$ are not chosen arbitrarily. The minimum values are selected by scanning them to obtain a convergent simulation case with the following selection priority: $\gamma$, $dx$, $\alpha_{fl,e}$ and $\beta$. Thus, the introduction of $\beta$ and $\gamma$ is somehow a compromise. According to the results in the previous section, the selected values for $\beta$ and $\gamma$ are relatively small and do not have significant influences on $\lambda_q$. The SOL normalized collisionalities at the separatrix are $\nu_e^* \sim \nu_i^* \sim 1$, which is slightly smaller than the value for the EAST case ($\nu_e^* \sim 3$). Although choosing a smaller $\alpha_{fl,e}$ that close to its lower limit (0.07) seems to be reasonable, we set $\alpha_{fl,e} = 0.3$ (this is a typical value that used in Refs. [27,28], with which Eq. (7) has a good approximation to the kinetic simulation) for faster saturation and avoidance of the convergency problem (it's difficult for the ITER simulation to converge for a smaller $\alpha_{fl,e}$). Since the parallel ion heat conduction is negligible and the lower limit for $\alpha_{fl,i}$ is 4.97, a relatively large $\alpha_{fl,i}$ is set, i.e., $\alpha_{fl,i} = 5$. The grid sizes in the radial and poloidal directions are 768 and 384, respectively, and the grid spacing is $0.41\rho_{s0}$ ($\rho_{s0} = 0.68$ mm) in both directions. The profile forcing region is set with $R_{PFR} = 0.5$. The ITER 15 MA baseline scenario conducts the deuterium (D) and tritium (T) fusion. Neglecting the fusion product (helium) and the impurities, we treat the main ions to be one species with $A = 2.5$ (50%/50% mixture of D and T) and $Z = 1$.

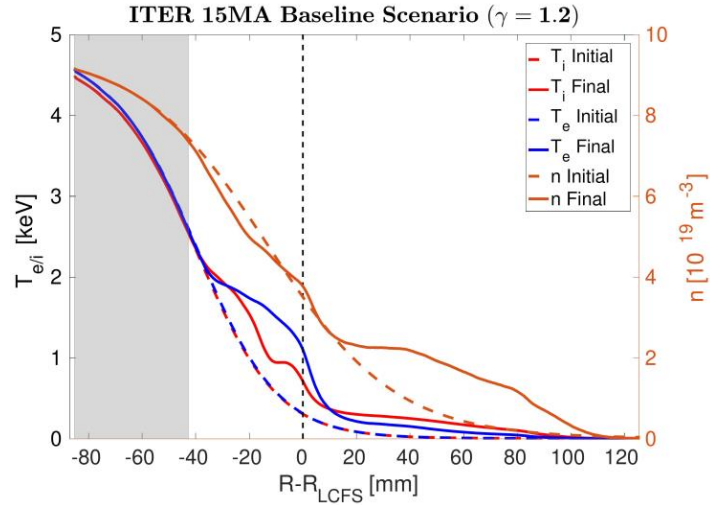

Figure 9 The initial (dashed lines) and final (solid lines) profiles of the density and electron and ion temperatures for the ITER 15 MA baseline scenario. The final profiles are averaged poloidally and temporally in the saturated phase. The profile forcing region is shaded with grey color ($R_{PFR} = 0.5$).

The solid lines in figure 9 represent the final profiles averaged poloidally and temporally in the saturated phase, which are elevated compared with the initial profiles (similar to the results produced by BOUT++ 6-field turbulence code in Ref. [16]), meaning that the given initial profiles generate significant transport of particles and heat into the SOL that cannot be efficiently removed by the parallel damping terms in the HESEL model mimicking the parallel losses in the SOL. Figure 10 shows a snapshot of the density in the saturated phase. It is clearly observed that the dynamical evolution is dominated by isolated revolving blob structures, which distribute in the SOL with their sizes decreasing along the radial direction towards the far SOL region and ultimately disappear near the wall region. Since the hyper-viscosity has not been applied to the SOL (the simulation fails to converge if the hyper-viscosity is turned on in the SOL) and a higher grid resolution requires more computation resources, the high-k numerical noise is not smoothed. As this noise does not change the blob structure and its dynamics, it can be neglected.

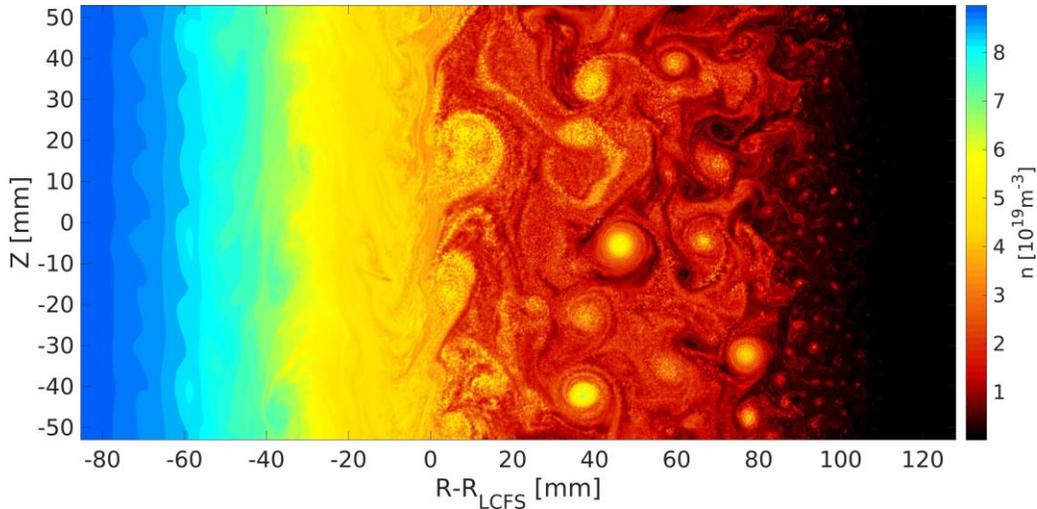

Figure 10 A snapshot of the density in the saturated phase for the simulation of the ITER 15 MA baseline scenario.

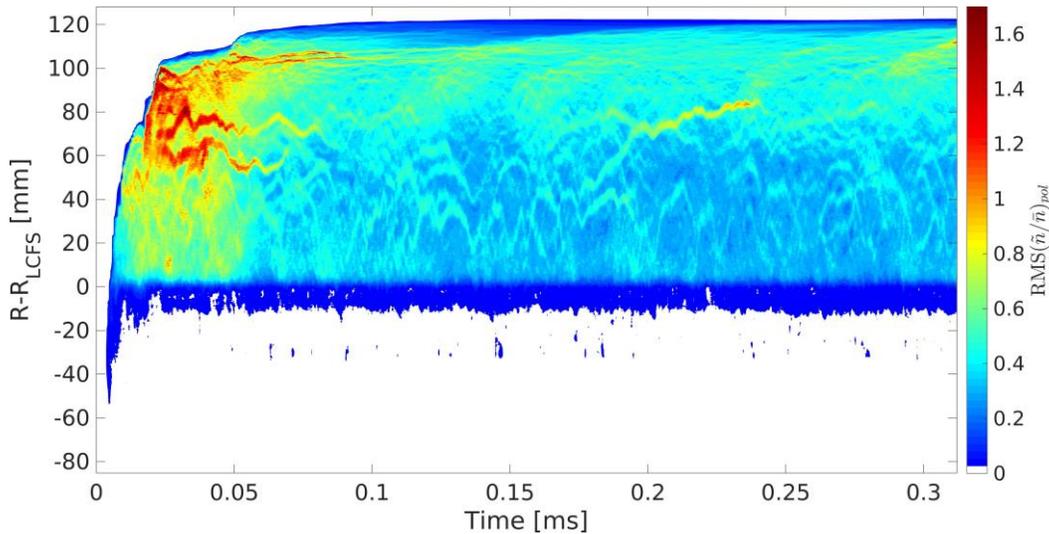

Figure 11 The contour plot of the root mean square (RMS) of the normalized density perturbation averaged in the poloidal direction with respect to time and radial location for the ITER simulation.

Figure 11 shows the contour of the root mean square (RMS) of the normalized density perturbation averaged in the poloidal direction, $\text{RMS}(\tilde{n}/\bar{n})_{pol}$, with respect to time and radial location. The $E \times B$ turbulent fluctuations in the edge region ($\text{RMS}(\tilde{n}/\bar{n})_{pol} \sim 0.1$) are much less violent compared with those in the SOL ($\text{RMS}(\tilde{n}/\bar{n})_{pol}$ is about 0.5 in the saturated phase and is greater than unity in the initial phase). This is different from the simulation results by BOUT++ 6-field turbulence code, where the largest turbulent fluctuation locates at the peak pedestal gradient inside the separatrix (see figure 7 in Ref. [16]), not in the SOL. But the fluctuation level inside the separatrix is similar, which might result from the similar pressure gradient of the initial profiles and the similar driving mechanism of the turbulence in the edge region.

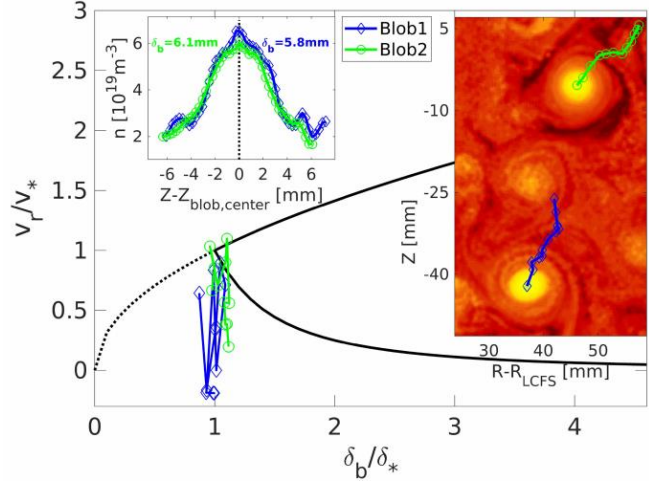

Figure 12 Characteristics of the selected blobs in the ITER simulation. The right inserted picture is trimmed from figure 10, where two blobs are selected with their trajectories being printed for a series of frames. The up-middle inserted picture shows the poloidal density profiles across their centers for these two blobs in the initial frame. The main frame compares the radial velocities of the selected blobs with the blob velocity scaling by the two-region model. The upper and lower solid black lines represent the resistive-ballooning and sheath-connected boundaries, respectively.

The tracers of the enhanced values in figure 11 can be interpreted as the radial travelling trajectories of the dense convective structures, i.e., the blobs. They are generated close to the separatrix, and generally travel radially outwards, but have a complicated trajectory and some of them turn around and move inward. This can be further confirmed by tracing the trajectories of the isolated blobs in figure 10 for a series of frames, as shown in figure 12. The inserted picture on the right side is trimmed and smoothed from figure 10 to show two isolated blob structures and their center of mass trajectories for several frames. Compared with the green trajectory (Blob2), the blue trajectory (Blob1) moves slower in the radial direction and even turns back radially in some frames. The up-middle inserted picture plots the poloidal profiles of these two blobs in the initial frame. These two blobs are not much elongated and have similar sizes (the full width at half maximum of the poloidal profile gives $\delta_b \approx 6$ mm) and masses. Known that the time interval for two adjacent frames is about 0.62 μs, the radial velocity of the blob, $v_r$, can then be estimated: $v_r$ ranges from -0.4 km/s to 2.2 km/s for Blob1 and ranges from 0.4 km/s to 2.7 km/s for Blob2. The main frame in figure 12 compares the blob velocity scaling with the well-known two-region model, where the blob size and velocity are normalized to $\delta_* = \rho_s^{4/5} L_c^{2/5} R^{-1/5}$ and $v_* = c_s(\delta_*/R)^{1/2}$, respectively [34]. The two-region model asserts that the velocity of the blob is related to the blob size and collisionality and is bounded by the upper resistive-ballooning (high collisionality) and the lower sheath-connected (lower collisionality) boundaries. For the traced blobs, their velocities are clearly bounded by the resistive-ballooning boundary, but most of them exceed the sheath-connected boundary. Considered that BOUT-HESEL utilizes the parameterized sheath boundary condition in the parallel direction [19] and blob theory or simulation studies use seeded blob [34-38], this violation may imply that the simplified two-region model is not sufficient to describe the blobs generated in turbulent simulations. Here, blobs are not purely isolated, may interact with other blob structures, and experience a strong poloidal flow often with a strongly sheared velocity.

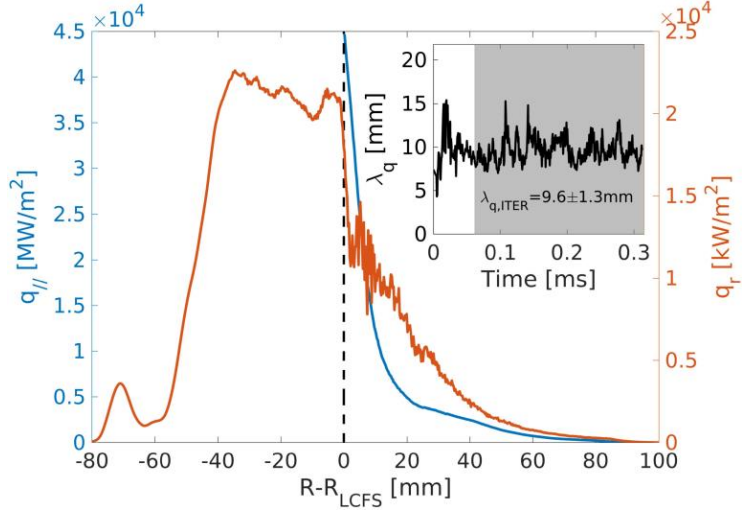

Figure 13 The radial profiles of the total parallel and radial heat fluxes for the ITER 15 MA baseline scenario. The inserted picture shows the evolution of $\lambda_q$, which gives $\lambda_{q,ITER}$ = 9.6±1.3 mm by averaging the data shaded with grey color.

The profiles of the total parallel and radial heat fluxes are shown in figure 13. The total power across the separatrix is estimated with Eq. (11) to be $P_{SOL}$ = 2.2 GW, which is about 17 times larger than the designed value in Ref. [33] ($P_{SOL} \approx$ 130 MW), but is nearly the same as the value in the simulation by BOUT++ 6-field turbulence code ($P_{SOL} \approx$ 2 GW) [16]. The elevations of the electron and ion temperatures (see figure 9) in the edge region contribute only ~3 out of 17 times for the difference of $P_{SOL}$. The dominant reason is the steep initial profiles, which generate significantly larger E × B turbulences in the edge region (the electron turbulent heat diffusivity $\chi_{e,turb}$ ranges from 29 m$^2$/s to 96 m$^2$/s). If the pressure gradient of the initial profiles is lower, $P_{SOL}$ will decrease (in the EAST simulations, $P_{SOL}$ decreases from ~0.8 MW to ~0.6 MW, when $\gamma$ increases from 1 to 1.2). For the parallel heat flux, the maximum value reaches up to 45 GW/m$^2$, which will reduce to ~20 GW/m$^2$ (close to the result by BOUT++ 6-field turbulence code and is one order of magnitude larger than the result by XGC1) if $\alpha_{fl,e}$ decreases to ~0.1. The evolution of $\lambda_q$ is inserted inside figure 13. $\lambda_q$ saturates quite fast and its mean value in the saturated phase (shaded with grey color) gives $\lambda_{q,ITER}$ = 9.6±1.3 mm, which is also close to the result produced by BOUT++ 6-field turbulence code (~11 mm), but is larger than the prediction by XGC1 (~6 mm). Remember that we use a relatively large $\alpha_{fl,e}$ and the prediction of ITER $\lambda_q$ will become even larger if a smaller $\alpha_{fl,e}$ is used. A quick estimate by the result shown in figure 6 gives $\lambda_{q,ITER} \approx 13$ mm if $\alpha_{fl,e}$ decreases to 0.15. Compared with the prediction by the Eich scaling (~1 mm), $\lambda_{q,ITER}$ simulated by BOUT-HESEL is also significantly larger.

## 5. Discussions on extrapolation from current devices to ITER

According to the simulation results of the EAST and ITER cases in this paper, BOUT-HESEL behaves similarly as XGC1 and BOUT++ 6-field turbulence code: comparisons of the simulated $\lambda_q$ with the Eich scaling give agreeable results for current devices, but predict significantly large $\lambda_q$ for the ITER 15 MA baseline scenario. To further understand this inconsistency, several simulations have been performed with parameters modified from the EAST (see section 3) and ITER (see section 4) cases. The main input and output parameters for these modified cases are listed in table 2 with the not changed parameters marked with NC.

Table 2. The main parameters of the HESEL simulations for the EAST, ITER, and their modified cases.

| | EAST | EAST-M1 | EAST-M2 | EAST-M3 | ITER | ITER-M1 | ITER-M2 |
|---|---|---|---|---|---|---|---|
| $R_c$ [m] | 1.85 | NC | 2.78 | 2.78 | 6.20 | 1.86 | NC |
| $a$ [m] | 0.43 | NC | 0.65 | 0.65 | 1.98 | 0.59 | NC |
| $B_t$ [T] | 2.44 | NC | NC | 3.66 | 5.53 | NC | NC |
| $B_p$ [T] | 0.21 | NC | NC | 0.32 | 1.33 | NC | NC |
| $q_{95}$ | 5.53 | NC | NC | NC | 3.60 | NC | NC |
| $n_0$ [$10^{19}$m$^{-3}$] | 1.84 | NC | NC | NC | 3.50 | 1.05 | 1.05 |
| $T_{e0}$ [eV] | 82 | NC | NC | NC | 310 | 93 | 93 |
| $T_{i0}$ [eV] | 83 | NC | NC | NC | 305 | 91 | 91 |
| $A$ | 2 | NC | NC | NC | 2.5 | NC | NC |
| $\alpha_{fl,e}$ | 0.6 | 0.3 | 0.3 | 0.3 | 0.3 | NC | NC |
| $\alpha_{fl,i}$ | 0.6 | 0.3 | 0.3 | 0.3 | 5 | NC | NC |
| $\beta$ | 0 | NC | NC | NC | 5 | NC | NC |
| $\gamma$ | 1 | NC | NC | NC | 1.2 | NC | NC |
| $dx$ [$\rho_{s0}$] | 0.76 | NC | NC | NC | 0.41 | NC | 0.51 |
| $\rho_{s0}$ [mm] | 0.66 | NC | NC | NC | 0.68 | 0.37 | 0.37 |
| $\nu_e^*$ | 2.85 | NC | NC | NC | 0.92 | 0.86 | 2.85 |
| $\chi_{e,turb}$ [m$^2$/s] | 5.9 | 7.5 | 11.1 | 6.4 | 25.5 | 6.3 | 50.9 |
| $v_{e,cd}$ [km/s] | 1288 | 879 | 893 | 733 | 2876 | 583 | 1442 |
| $\lambda_{q,Eich}$ [mm] | 4.10 | 4.10 | 4.10 | 2.53 | 0.45 | 0.45 | 0.45 |
| $\lambda_{q,HD}$ [mm] | 4.19 | 4.19 | 4.19 | 2.79 | 1.72 | 0.94 | 0.94 |
| $\lambda_{q,HESEL}$ [mm] | 4.25 | 5.07 | 7.93 | 7.11 | 9.59 | 5.51 | 16.70 |
| $\lambda_{q,HESEL}^{\#}$ [mm] | 5.04 | 5.04 | 7.87 | 6.97 | 9.30 | 2.47 | 9.30 |

Note: The parameters that are not changed are marked with NC for the EAST and ITER modified cases. The values of $\chi_{e,turb}$ and $v_{e,cd}$ are averaged in the range of $R$-$R_{LCFS} \in [0, \lambda_q]$. $\lambda_{q,HESEL}^{\#}$ is evaluated by Eq. (14).

The ITER-M1 case shrinks the machine size and the plasma profiles by 70%, approaching similar levels as the EAST case, accordingly $\lambda_{q,HESEL}$ (5.5 mm) is reduced to the same level as for the EAST-M1 case ($\lambda_{q,HESEL}$ = 5.1 mm). To examine whether the machine size or the plasma profiles account for the decrease of $\lambda_{q,HESEL}$, the machine size is restored to the ITER case and the plasma profiles are kept shrunk as for the ITER-M1 case, this is represented by the ITER-M2 case. $\lambda_{q,HESEL}$ for this case (16.7 mm) is even larger than that for the ITER case (9.6 mm), suggesting that the decrease of the machine size is the main reason for the decrease of $\lambda_{q,HESEL}$ from the ITER case to the ITER-M1 case. It is well known that $\lambda_q$ is determined by the competition between the radial and parallel heat transports in the near SOL. In our simulations, the radial heat transport is dominated by the electron and ion E × B turbulences, and the parallel heat transport is dominated by the electron conduction, as shown in figure 6. The radial electron and ion turbulent heat transports can be characterized by the electron ($\chi_{e,turb}$) and ion ($\chi_{i,turb}$) turbulent heat diffusivities, respectively. The parallel electron heat conduction can be characterized by the damping velocity of electron conduction, $v_{e,cd} \equiv q_{cond,e}/p_e$. Assume that the electron and ion temperature gradients are similar and $\chi_{e,turb}/\chi_{i,turb}$ is a constant, then the total radial heat diffusivity, $\chi_{e,turb} + \chi_{i,turb}$, is proportional to $\chi_{e,turb}$. If $\lambda_q/\lambda_T$ is assumed to be a constant ($\lambda_T$ is the SOL temperature width), $\lambda_q \propto \sqrt{L_c \chi_{e,turb}/v_{e,cd}}$ according to Eq. (21) in Ref. [19]. Therefore, we can argue that $L_c$, $\chi_{e,turb}$ and $1/v_{e,cd}$ are in the same order on determining $\lambda_q$. From the ITER-M1 case to the ITER-M2 case where the machine size is increased, $L_c$, $\chi_{e,turb}$ and $v_{e,cd}$ increase to about 3.3, 8.1 and 2.5 times, respectively (about $\sqrt{3.3 \times 8.1/2.5} \approx 3.3$ times in total, consistent with the increase of $\lambda_{q,HESEL}$), indicating that the increase of the machine size mainly enhances the radial turbulent transports. From the ITER-M2 case to the ITER case where the plasma profiles are elevated, $\chi_{e,turb}$ and $1/v_{e,cd}$ both decrease by about 50% (about 0.5 times in total), which cannot cancel the positive effect of the increase of the machine size from the ITER-M1 case to the ITER case.

For ITER and its modified cases in table 2, the predictions of $\lambda_q$ by the Eich scaling stay the same, not responding to the change of the plasma profiles or machine size, which is different from the results given by the BOUT-HESEL code. Since the consistency still holds for the EAST case, the machine size and the toroidal magnetic field are increased stepwise by 50%, which are represented by the EAST-M2 and EAST-M3 cases, respectively. When the machine size is increased (from the EAST-M1 case to the EAST-M2 case), $v_{e,cd}$ does not change much but $L_c$ and $\chi_{e,turb}$ increase by 50% and 48%, respectively, resulting in a 56% increase of $\lambda_{q,HESEL}$. Figure 14 shows the snapshots of the density for the EAST-M1 (top panel), EAST-M2 (middle panel), and EAST-M3 (bottom panel) cases in the saturated phase. We see that the number and the size of the turbulent structures increase and the fluctuations become more violent, when the machine size is increased. When $B_t$ is increased (from the EAST-M2 case to the EAST-M3 case), $\chi_{e,turb}$ and $v_{e,cd}$ both decrease by about 43% and 18%, respectively, resulting in a 10% decrease of $\lambda_{q,HESEL}$. According to figure 14, the size of the turbulent structures becomes smaller with the number increasing. To sum up, the results obtained from the EAST and ITER modified cases: the increase of the machine size strengthens the radial turbulent transports and slows down the parallel heat transports ($L_c$ is increased); the increase of $B_t$ reduces the radial turbulent transports and weakens the parallel heat conduction. These lead to a strong positive scaling dependence of $\lambda_q$ on $R_c$ and a weak negative scaling dependence of $\lambda_q$ on $B_t$, respectively.

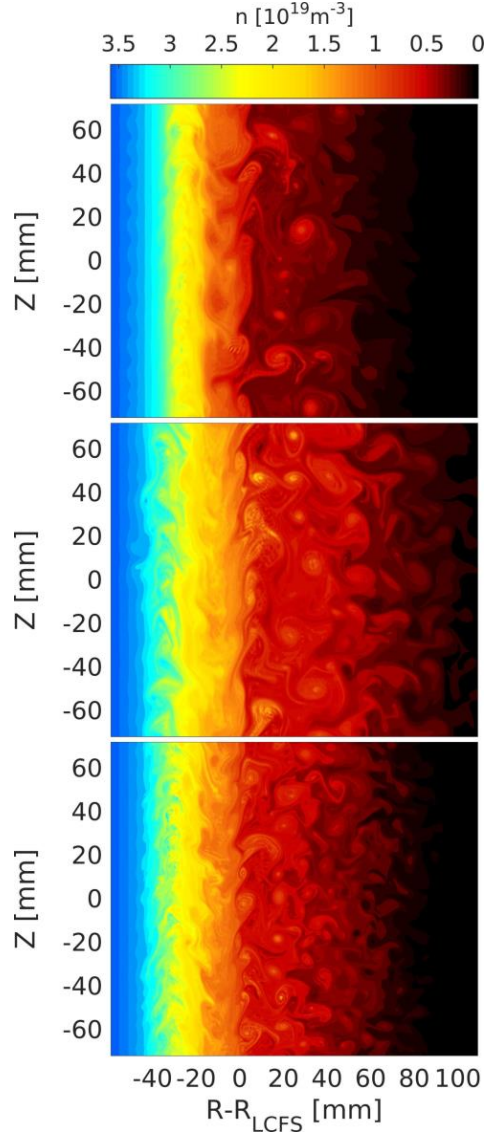

Figure 14 The snapshots of the density for the EAST-M1 (top panel), EAST-M2 (middle panel, the machine size is increased by 50% compared with the EAST-M1 case), and EAST-M3 (bottom panel, $B_t$ is increased by 50% compared with the EAST-M2 case) cases in the saturated phase.

In table 2, $\lambda_{q,Eich}$ only responds to the change of $B_t$ due to the change of $B_p$ ($B_p$ is calculated with the relation, $q_{95} \sim q_{cyl} \approx \frac{aB_t}{R_c B_p}$). For all devices in the multi-machine database (the basic parameters are listed in table 3) in Ref. [3], the Eich scaling is

$$\lambda_{q,Eich} = 0.63 B_p^{-1.19}, \qquad (12)$$

which has no scaling dependence on $B_t$. However, this scaling dependence appears if the spherical tokamaks (NSTX and MAST) are excluded and $B_p$ is replaced by $q_{95}$ [3], i.e.,

$$\lambda_{q,Eich}^{\#} = 0.70 R_c^0 B_t^{-0.77} q_{95}^{1.05} P_{SOL}^{0.09}, \qquad (13)$$

where $P_{SOL}$ is the experimental measurement, not the simulation estimation by Eq. (11). For the BOUT-HESEL code, the previous L-mode scaling of $\lambda_q$ gives, $\lambda_q = 0.26 R_c^{1.32} B_t^{-0.33} q_{95}^{1.30}$ [19], which shows different scaling dependences on $R_c$ and $B_t$. A quick estimation of HESEL H-mode scaling can be obtained by using parameters in table 2. By comparing $\lambda_{q,HESEL}$ and $B_t$ for the EAST-M2 and EAST-M3 cases, the scaling exponent of $B_t$ equals to log(7.11/7.93)/log(3.66/2.44) ≈ -0.27, which is close to that in the HESEL L-mode scaling. Here the exponent of $B_t$ is set to -0.3 if the L-mode scaling is considered. For the scaling exponent of $R_c$, the EAST and ITER modified cases give 1.10 and 0.92, respectively. Together with the scaling exponent of $R_c$ in the L-mode scaling, the mean value 1.1 is chosen. As discussed in Ref. [19], the scaling dependences of $\lambda_q$ on $q_{95}$ and $R_c$ are the scaling dependence on the parallel ballooning length $L_b \approx q_{95} R_c$. So, the same scaling exponent of $q_{95}$ is chosen as that of $R_c$, which is reasonable according to the numerical L-mode scalings obtained in Ref. [19] and in figure 1(b). After substituting the parameters of the EAST-M1 case (to get the scaling amplitude), the estimated HESEL H-mode scaling of $\lambda_q$ is

$$\lambda_{q,HESEL}^{\#} = 0.51 R_c^{1.1} B_t^{-0.3} q_{95}^{1.1}. \qquad (14)$$

With no surprise, $\lambda_q$ evaluated by the above scaling agrees well with the simulated $\lambda_q$ for the EAST modified cases, as listed in table 2. The surprise is that its prediction for the ITER case also shows close agreement with an error of -6.8%. The inconsistency between $\lambda_{q,HESEL}$ and $\lambda_{q,HESEL}^{\#}$ in table 2 for the ITER modified cases arises from the absence of the scaling dependence on the plasma temperature. According to the previous studies in Refs. [8,19], $\lambda_q$ is proportional to $T_{LCFS}^{-0.5}$. This inconsistency can then be covered: a decrease of $T_{LCFS}$ by 70% from the ITER case increases $\lambda_q$ by $0.3^{-0.5} - 1 \approx 83\%$.

In table 3, the different scaling dependences on $R_c$ and $B_t$ between Eq. (13) and Eq. (14) are compared based on parameters in the multi-machine scaling database [3]. For the $\lambda_{q,Eich}^{\#}$ scaling, $R_c^0 B_t^{-0.8} \sim 0.5$ covers all current devices except for the spherical tokamaks, and for the $\lambda_{q,HESEL}^{\#}$ scaling, $R_c^{1.1} B_t^{-0.3} \sim 1.5$ can cover all current devices except for C-Mod. Note that $R_c^0 B_t^{-0.8}$ for C-Mod is about 50% smaller than $R_c^0 B_t^{-0.8} \sim 0.5$. Then the $\lambda_{q,HESEL}^{\#}$ scaling can also be covered by the doubled value of $R_c^{1.1} B_t^{-0.3}$ for C-Mod. Since $R_c^{1.1} B_t^{-0.3}$ has similar values (the value for C-Mod is doubled) for all the devices in the multi-machine scaling database, its scaling significance is weakened in the regressions. This probably explains why the Eich scaling in Eq. (12) has no scaling dependence on neither $R_c$ nor $B_t$. When it relates to ITER, the significance of $R_c^{1.1} B_t^{-0.3}$ comes out, showing a ~3 times larger value than that for the current devices. When extrapolating the scalings to the ITER 15 MA baseline scenario, the scaling difference ($R_c^{1.1}B_t^{0.5}$) between the Eich scaling and the HESEL H-mode scaling rises up to 17.5/3 ≈ 6 times. This could explain why current codes (XGC1, BOUT++ 6-field turbulence code, and BOUT-HESEL) can produce consistent results with the Eich scaling for current devices, but predicts a much larger $\lambda_q$ for ITER.

Table 3. The basic parameters and scaling dependences of $\lambda_q$ on $R_c$ and $B_t$ for current devices and ITER

|  | JET | DIII-D | AUG | C-Mod | NSTX | MAST | EAST | ITER |
|---|---|---|---|---|---|---|---|---|
| $R_c$ [m] | 2.95 | 1.74 | 1.65 | 0.7 | 0.87 | 0.87 | 1.85 | 6.20 |
| $a/R_c$ | 0.32 | 0.29 | 0.31 | 0.31 | 0.69 | 0.70 | 0.24 | 0.32 |
| $B_t$ [T] | 1.1–3.2 | 1.2–2.2 | 1.9–2.4 | 4.6–6.2 | 0.4–0.5 | 0.4 | 2.44 | 5.53 |
| $B_p$ [T] | 0.2–0.7 | 0.2–0.5 | 0.2–0.5 | 0.5–0.8 | 0.2–0.3 | 0.1–0.2 | 0.21 | 1.33 |
| $q_{95}$ | 2.6–5.5 | 3.2–7.3 | 2.6–5.1 | 3.8–6.6 | 5.5–9.0 | 4.9–6.8 | 5.53 | 3.60 |
| $R_c^0 B_t^{-0.8}$ ($\lambda_{q,Eich}^{\#}$) | 0.9–0.4 | 0.9–0.5 | 0.6–0.5 | 0.3–0.2 | 2.1–1.7 | 2.1 | 0.49 | 0.25 |
| $R_c^{1.1} B_t^{-0.3}$ ($\lambda_{q,HESEL}^{\#}$) | 3.2–2.3 | 1.7–1.5 | 1.4–1.3 | ~0.4 | ~1.1 | 1.1 | 1.51 | 4.45 |
| $R_c^{1.1} B_t^{0.5}$ | 3.4–5.9 | 2.0–2.7 | 2.4–2.7 | 1.4–1.7 | 0.5–0.6 | 0.5 | 3.07 | 17.50 |

Note: Except for the EAST and ITER cases in this paper, the parameters for the other devices are picked from table 1 of the multi-machine scaling database in Ref. [3].

As mentioned in the introduction, the drift-turbulence competition theory [16] offers another explanation for the prediction difference of ITER $\lambda_q$ between the simulations and the Eich scaling. In section 3, we also see a good agreement on the prediction of EAST $\lambda_q$ between the HESEL simulation and the HD model [4] (see figure 8). Since the magnetic $B \times \nabla B$ drift plays the dominant role on determining $\lambda_q$ in both the drift-turbulence competition theory and the HD model, it is reasonable to evaluate its influence on $\lambda_q$ in HESEL simulations. However, the $B \times \nabla B$ drift in the HESEL model is in the poloidal direction (HESEL is a 2D model, see Ref. [19]), meaning that magnetic drifts have no contribution to the radial particle and heat transports. To estimate the effect of $B \times \nabla B$ drift on the radial particle/heat transports in the SOL, we introduce the equivalent radial magnetic drift (ERMD). In a LSN configured plasma with ion $B \times \nabla B$ drift direction pointing downwards, the ion ERMD velocity ($v_{r,ERMD}$) can be obtained by calculating the mean value of the radial component of the ion $B \times \nabla B$ drift velocity from the OMP to the lower X-point over the same flux surface. Assume that the cross-section of the plasma is a circle, $v_{r,ERMD}$ is written as $v_{r,ERMD} = \frac{2}{\pi}\int_0^{\frac{\pi}{2}} v_{B\times\nabla B} \sin\theta d\theta$, where $\theta$ is the poloidal angle deviating from the OMP ($\theta$ is $\frac{\pi}{2}$ at the lower X-point). Using the poloidal mean of the ion temperature on the OMP in the HESEL model to approximate $v_{B\times\nabla B}$ during the drifting process, $v_{r,ERMD}$ is simplified to

$$v_{r,\text{ERMD}}(r) = \frac{4\bar{T}_i(r)}{\pi Z e R_c (R_c + a) B_0}(R_c + 0.5a + 0.5r), \qquad (15)$$

where $r \equiv R-R_{LCFS} \geq 0$. The HD model assumes that the magnetic drift dominates the cross-field transport of particles into the SOL and the parallel flow velocity is half of the ion sound speed [4]. Then, $\lambda_q$ can be approximated by Eq. (15) as $\lambda_q \approx v_{r,ERMD}(0)\frac{\pi}{2}q_{cyl}R_c/(0.5c_s) \approx 4\bar{T}_{i,LCFS}a/(ZeB_pc_sR_c)$, which is same as the $\lambda_q$ scaling given by the HD model (see the introduction).

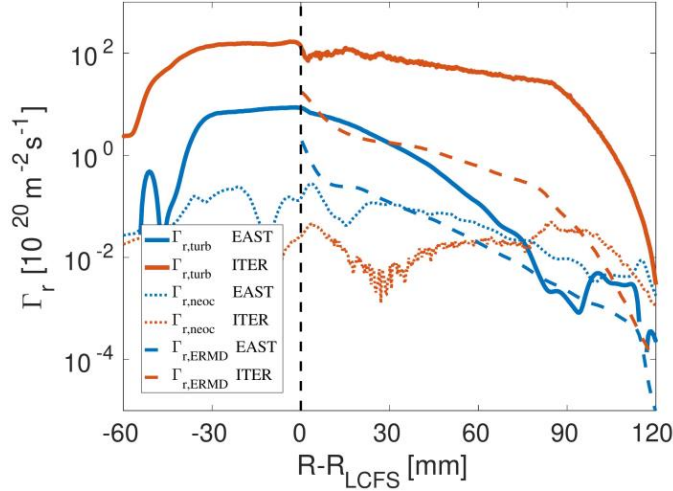

Figure 15 The distributions of radial particle flux ($\Gamma_r$) for E × B turbulence (turb, solid lines), neo-classical diffusion (neoc, dotted lines) and ion equivalent radial magnetic drift (ERMD, dashed lines), calculated from the EAST and ITER cases. The vertical dashed line represents the separatrix.

With Eq. (15), we can evaluate the radial particle flux resulted from the ion magnetic drift by $\Gamma_{r,ERMD} \equiv nv_{r,ERMD}$. $\Gamma_{r,ERMD}$ is compared with the radial turbulent ($\Gamma_{r,turb}$) and neo-classical ($\Gamma_{r,neoc}$) particle fluxes for the EAST and ITER cases in figure 15. For both cases, $\Gamma_{r,ERMD}$ is about one order of magnitude smaller than $\Gamma_{r,turb}$ in the near SOL ($\Gamma_{r,neoc}$ is nearly negligible). This means that the magnetic drift is not the determinant mechanism for cross-field transports if it is included through ERMD in the HESEL simulations and we can explain the prediction difference of ITER $\lambda_q$ between the simulations and the Eich scaling, where turbulent transport is the dominant mechanism for the radial heat transport into and across the SOL. Since the fact that turbulence dominates the radial heat transport for ITER applies to the simulations by all three codes (XGC1, BOUT++ 6-field turbulence code, and BOUT-HESEL) and magnetic drifts decrease with $R_c$ and $B_t$ increasing (they are weakened for ITER), the major difference between our explanation and the drift-turbulence competition theory is in which mechanism dominates the radial particle/heat transports in the SOL for current devices. Except for the results shown in this paper, many other simulation works [12,14,39,40] show the radial turbulent transport as the dominant player in setting the SOL power/particle widths for current devices. Compared with the experimental scalings [6,8], the turbulence-based scalings in Refs. [18,19] show consistent results on the negative scaling dependence of $\lambda_q$ on the edge plasma temperature. While the drift-based scaling gives an opposite scaling dependence ($\lambda_q \propto T_{LCFS}/c_s \propto T_{LCFS}^{1/2}$). The inconsistency between the turbulence-based scalings and the Eich scaling is the scaling dependence of $\lambda_q$ on the machine size (the drift-based scaling shows consistency), which appears to be explained by the cancellation of scaling significances between $R_c$ and $B_t$ in the Eich scaling database. Although it appears that the turbulence-based explanation shows more consistent results with the previous simulation and experimental studies than the drift-based explanation, there still needs more evidences from the experimental studies and the reason why the simple drift-based scaling agrees with the Eich scaling needs further investigations.

## 6. Summaries

In this paper, the newly upgraded BOUT-HESEL code is used to simulate H-mode plasmas with low collisionality for the first time. The validation against the previous implementation using EAST L-mode discharges produces results more consistent with the Eich scaling. The obtained numerical L-mode scaling of $\lambda_q$ shows good agreements with the multi-machine L-mode scaling and the Eich H-mode scaling. A typical EAST H-mode discharge is selected to further validate the HESEL model against the experiment and to provide essential information for simulating the ITER 15 MA baseline scenario. It is found that the electron flux-limited coefficient $\alpha_{fl,e}$ influences the parallel heat conduction (dominates the parallel heat transport) significantly in the SOL (the radial heat flux in the edge region is not influenced), making the simulated $\lambda_q$ sensitive to the value of $\alpha_{fl,e}$. The comparison of the PDF of the parallel particle flux with measurements by reciprocating probes sets $\alpha_{fl,e} = 0.6$ for the EAST simulation. The power entered into the SOL in the simulation is in the same order of magnitude as the total heating power in the experiment and the simulated $\lambda_q$ is nearly the same as the predictions by the Eich scaling and the heuristic drift-based model.

Based on the EAST simulation, the level of hyper-viscosity (applied only in the edge region) and stretch of the initial profiles (changes the pressure gradients with the values at the separatrix fixed) are scanned, confirming that they have limited influences on $\lambda_q$ for relatively low values. Then, the hyper-viscosity and the stretch of the initial profiles with moderate levels are applied to the simulation of the ITER 15 MA baseline scenario (Q = 10) developed by Kim *et al.* [33] to handle the high-k components of the spatial spectrum in the numerical scheme with finite grid resolution. The ITER simulation gives $\lambda_{q,ITER} = 9.6$ mm, which is much larger than the prediction by the Eich scaling (~1 mm), but communicates with the simulation results by XGC1 (~6 mm) and BOUT++ 6-field turbulence code (~11 mm). The ITER simulation shows that the radial particle/heat transports in the SOL are dominated by blobby transports and the maximum level of density fluctuation locates in the SOL (the RMS of the normalized fluctuation reaches up to ~0.5, which is about 5 times larger than that in the edge region). The typical blob size is about 6 mm and the two-region model cannot fully explain the scaling of blob velocity for the ITER simulation.

A few cases modified from the EAST and ITER cases are used to understand the difference of extrapolation of $\lambda_q$ from current devices to ITER between the Eich scaling and the numerical simulations. Combining the HESEL L-mode scaling and the simulation results of the EAST modified cases, the HESEL H-mode scaling is estimated to be $\lambda_{q,HESEL}^{\#} = 0.51 R_c^{1.1} B_t^{-0.3} q_{95}^{1.1}$. This scaling predicts a surprisingly consistent $\lambda_{q,ITER}$ (9.3 mm) with that for the ITER case ($\lambda_{q,ITER}$ = 9.6 mm). Compared with the Eich scaling with the spherical tokamaks excluded, i.e., $\lambda_{q,Eich}^{\#} = 0.70 R_c^0 B_t^{-0.77} q_{95}^{1.05} P_{SOL}^{0.09}$, $\lambda_{q,HESEL}^{\#}$ has a strong positive scaling dependence on the major radius $R_c$ and a weaker negative scaling dependence on the toroidal magnetic field $B_t$. The increase of $R_c$ not only slows down the parallel heat transports by increasing the parallel connection length, but also strengthens the radial turbulent transports significantly. While the increase of $B_t$ weakens both the radial turbulent transports and the parallel heat transports, but the former effect is stronger. Further investigation of this scaling difference based on the multi-machine database used for the Eich scalings reveals that the regression significance of $R_c$ appears to be shaded by that of $B_t$ for current devices — but not for ITER — in the multi-machine scaling database. This might explain why the Eich scaling with all devices included, i.e., $\lambda_{q,Eich} = 0.63 B_p^{-1.19}$, has no scaling dependences on $R_c$ and $B_t$ and why simulation codes (like XGC1, BOUT++ 6-field turbulence code, and BOUT-HESEL) can reproduce the Eich scaling for current devices, but predict a much larger $\lambda_q$ for ITER. This explanation differs from the drift-turbulence competition theory [16], where magnetic drift is considered as the key physical mechanism, but found to be trivial for both the EAST and ITER simulations. The strong positive scaling dependence of $\lambda_q$ on $R_c$ in the HESEL H-mode scaling should be beneficial for solving the power exhaust issue for large size fusion-grade magnetically confined devices.

## Acknowledgments

The author X. Liu thanks Dr. Ning Yan, Dr. Xiang Han, Mr. Tianfu Zhou, Mr. Shengyu Fu, and Mr. Yifei Jin for providing EAST experimental data, and appreciates the help from Dr. Sun Hee Kim, Dr. Zeyu Li and Dr. Xueqiao Xu for sharing plasma profiles of the ITER 15 MA baseline scenario. This work was supported by the Natural Science Foundation of China (Nos. 12005260, 11922513), and the National Key Research and Development Program (No. 2017YFE0301300). This work was also partially supported by Institute of Energy, Hefei Comprehensive National Science Center under Grant No. GXXT-2020-004. Numerical computations were performed on the ShenMa High Performance Computing Cluster in Institute of Plasma Physics, Chinese Academy of Sciences.